\documentclass[]{interact}

\usepackage{epstopdf}
\usepackage[caption=false]{subfig}
\usepackage[subfigure]{tocloft}

\usepackage[numbers,sort&compress]{natbib}
\bibpunct[, ]{[}{]}{,}{n}{,}{,}
\usepackage[utf8]{inputenc}
\usepackage[english]{babel}                
\usepackage[T1]{fontenc}                    
\usepackage{ragged2e,anyfontsize}
\DeclareUnicodeCharacter{1F30}{$\iota$}

\usepackage{lineno,hyperref}
\usepackage{natbib}
\usepackage{adjustbox}
\usepackage{longtable}
\usepackage{booktabs}
\usepackage{xcolor}
\usepackage{caption}
\usepackage{rotating}
\usepackage{lscape}
\usepackage{booktabs}
\usepackage{soul}
\usepackage{textcomp}
\usepackage{multirow}
\usepackage{multicol}
\usepackage{tabu}
\usepackage{tocloft}
\usepackage{blindtext}
\usepackage{mathtools}
\usepackage{blkarray}
\usepackage{subcaption}
\usepackage{float}
\usepackage{caption}
\usepackage{rotating}
\usepackage{lscape}
\usepackage{array}
\usepackage{booktabs}
\usepackage{soul}
\usepackage{textcomp}
\usepackage{amsmath}
\usepackage{amssymb}
\usepackage{setspace}
\usepackage{bm}
\usepackage{makecell}

\theoremstyle{plain}

\theoremstyle{definition}

\theoremstyle{remark}

\begin{document}

\title{Text Data Analysis of Maternal Narratives: Albanian Women in Italy}

\author{
\name{Eleonora Miaci\textsuperscript{a}\thanks{CONTACT Eleonora Miaci. Email: eleonora.miaci@unimi.it} and Emiliano Seri\textsuperscript{b}}
\affil{\textsuperscript{a}Department of Social and Political Sciences, University of Milan, Milan, Italy; \textsuperscript{b}Department of Enterprise Engineering “Mario Lucertini”, University of Rome Tor Vergata, Rome, Italy}
}

\maketitle

\begin{abstract}
Despite growing interest in migration studies, research on motherhood among migrant women in Italy remains limited. This study contributes to the literature by examining the family trajectories of Albanian women in Italy, exploring how their migration patterns and experiences have shaped these life aspects.
We conducted a comprehensive textual analysis to find the main topics of 30 semi-structured interviews with Albanian mothers living in Milan, Rome, and Bari. After pre-processing the text, we performed an exploratory analysis to identify key features and explore word relationships. The predominant dimensions that emerged relate to family management, work paths and schedules, and strategies and concerns arising from the trade-off between work and childcare. Subsequently, we stratified the sample by entry channel into Italy (study and work, reunification, and irregular channel) and applied Latent Dirichlet Allocation to model each sub-corpus as a mixture of topics.
Our results resonate with existing literature \citep{ortensi2015engendering} on the key role of female migratory patterns in shaping post-migration fertility. Interviewees who entered Italy through various migratory channels not only differ in their characteristics and migration experiences but also exhibit dissimilar fertility desires and behaviors, motherhood trajectories, and conceptions of their role as mothers and family ideals. These differences influence their priorities and level of commitment to family and work obligations.
\end{abstract}


\section{Introduction}
\label{sec:intro}

In a country characterized by lowest-low fertility like Italy \citep{kohler2002emergence}, migrant residents \footnote{In this paragraph, the term "migrants" has been predominantly used, as it is more neutral and inclusive, avoiding the potentially divisive connotations of the term "foreigners." However, it is important to clarify that all the statistics cited refer specifically to the population with foreign citizenship.}, who are on average younger than the Italian female population, have contributed to slowing the decline in births \citep{istat2021, mussino2012fertility}. This contribution is due to their significant presence in the fertile age group, a younger average age at childbirth (29.6 years), and a total fertility rate (TFR) close to the replacement level (1.8 children per woman in 2022) \citep{istat2022}. Despite this, the fertility differential between migrant and native women in Italy has decreased significantly over the years. While the gap between the fertility rate (1.18 in 2022) and the mean age at childbirth of native women (32.9 in 2022) remains considerable, the reduction in TFR experienced by migrants over the past decade has been sharper (-14\% for foreign women compared to -9\% for Italian women since 2012). Additionally, the increase in the mean age at childbirth has been more pronounced for migrant women (+4.6\% compared to +2.8\% for native women since 2012).
Among the five largest migrant communities in Italy (Romanian, Albanian, Chinese, Moroccan, Ukrainian), significant differences in reproductive behaviors emerge. Notably, the fertility rates of Moroccan and Albanian women are relatively high, and their modal age at childbirth is younger compared to both other resident communities in Italy \citep{miaci2023}.

Building on these macro-data premises, 30 in-depth interviews with Albanian mothers residing in Italy were conducted to delve deeper into expectations, representations, and practices related to motherhood and family life. While we acknowledge that such a small sample cannot represent the entire Albanian female population in Italy, it is beyond the scope of this paper to generalize our findings to this broader group. Furthermore, several studies have focused on revealing how the fertility of Albanian women in Italy is mainly explained by adaptation \citep{garcia2023comparison, impicciatore2020migrants} and how the younger age structure of this population results in relatively high fertility rates \citep{miaci2023}.

In this paper, we aim to explore the narrative modes through which these women reconstruct their motherhood trajectories and to deepen our understanding of how the social conditions they face, their individual characteristics, and their migratory paths affect their family life.

We quantitatively evaluate the corpus derived from the interview material through statistical text analysis, delving into the topics structures of the interviews. 
We stratify the sample based on the “reason of migration” variable (study and work, reunification, and irregular channel) and we apply a Latent Dirichlet Allocation (LDA) \citep{blei2003latent} to identify four main topics and model each subcorpus as a mixture of these topics. A model-based approach such as LDA allows the analysis to benefit from the inferential framework of statistics, addressing practical questions arising during clustering, such as determining the number of clusters and detecting and treating outliers \citep{Bouveyron2019Jun}. Moreover, in the LDA “each word is modeled as a finite mixture over an underlying set of topics. Each topic is, in turn, modeled as an infinite mixture over an underlying set of topic probabilities \citep{blei2003latent}”. This approach enables a deep understanding of the nuances in word usage and how words acquire different meanings in relation to the topics they are used in.

This analysis revealed that, in line with existing literature on the subject, it is important to include the migratory pattern as a variable that mediates reproductive and family patterns after migration \citep{ortensi2015engendering}. 
In the subcorpus of interviewees who entered Italy through family reunification, we found a prioritization of the establishment and care of their family. In contrast, among employment seekers, there was a greater commitment to their jobs and the attempt to balance family with their professional careers, in line with the findings of \cite{nedoluzhko2007migration} and \cite{mussino2012fertility}. 
Notably, both groups exhibited evidence of maintaining transnational connections with their country of origin, manifested through frequent cross-border visits, sustained communication with familial networks, remittance flows, or considerations of return migration. 
Conversely,  migrants entered in Italy with irregular immigration status were characterized by a heightened focus on the challenges and uncertainties faced during their initial settlement, particularly regarding employment. Their narratives reflected also a shift from reactive to proactive strategies, evolving from immediate survival approaches to longer-term plans for stability and socioeconomic advancement \citep{triandafyllidou2020irregular}.  

The article is structured as follows. Section \ref{Sec:background} presents a general theoretical framework of the existing theories and literature on migrants’ fertility and motherhood and the Albanian contribution to fertility in Italy. Section \ref{Sec:data} details the corpus of data, including a description of the pre-processing operations employed to prepare it for analysis. Section \ref{Sec:method} explains LDA methodology for topic modeling. Section \ref{Sec:application} presents the applications and results of this analysis, with one paragraph covering exploratory results and another discussing the results from LDA. The article concludes with a discussion section \ref{Sec:conclusion} that summarizes the main findings, implications, suggestions for future research, and conclusive remarks.

\section{Background and Literature Review}
\label{Sec:background}
\subsection{Theories and Literature on Migrants’ Fertility and Motherhood}

The international literature consistently highlights significant fertility differentials between foreign-born and native-born populations across most high-income countries \citep{wilson2020understanding}. While trends vary by country and migrant group, migrant women typically exhibit higher fertility rates than native women, though a gradual convergence in fertility levels often occurs over time \citep{ sobotka2008overview}.

To explain the differential fertility of migrant women, the literature presents several hypotheses regarding the impact of new social environments on reproductive behaviour (for a review see  \citep{kulu2005migration}. These theories are not mutually exclusive and may coexist, with varying relevance depending on the specific context. Some theories emphasize contextual influences, either from the country of origin (socialization) or the destination country (cultural entrenchment, adaptation). Selection theory posits that migrants represent a group whose fertility preferences align more closely with the destination population than with their origin country.
Other theories examine fertility from a longitudinal perspective, situating it within the individual's life course \citep{wilson2020understanding}. These approaches focus on how migration interrupts the reproductive pathway (disruption) or interrelates with it in a series of reciprocal influences over time (interrelation) \citep{andersson2004childbearing, gonzalez2016partnership, mulder1993migration, milewski2009fertility}.

Migration motives emerge as a key predictor of migrant reproductive behaviors \citep{ortensi2015engendering}. The literature exploring the relationship between gender, migration patterns, and fertility highlights the diverse characteristics and experiences of migrants, revealing varying outcomes in fertility and family formation depending on the primary reason for migration \citep{kulu2005migration, nedoluzhko2007migration, mussino2012fertility, tonnessen2020fertility}. 
Women migrating for family reunification tend to align their reproductive behaviors with the goal of maintaining and strengthening family ties, often resulting in higher fertility rates. In contrast, employment seekers tend to prioritize work over family formation, potentially delaying childbearing until achieving financial stability \citep{ortensi2015engendering}. Morevoer, primary migrants may struggle to balance work and family life, especially in the absence of immediate family support, in contrast, those who migrate for family reunification may experience a smoother transition into family life but could face economic challenges if their focus on family care limits their engagement in the labor market. The differing priorities between these two groups can lead to distinct social dynamics and integration experiences in the host country \citep{ortensi2015engendering}.

Extensive research has examined the interplay between migration, reproductive behaviors, and family structures \citep{oropesa1997immigrant, toulemon2004fertility, cooke2008migration, milewski2009fertility}.
Women's participation in international migration is significantly shaped by gender roles and norms from their countries of origin, resulting in diverse outcomes in destination countries \citep{carling2005gender, hiller2007reassessing}.
Migration involves a reconfiguration of value systems and family relationships, impacting gender dynamics within conjugal couples and intergenerational ties with families and communities of origin. 
For both parents and children, migration often triggers a redefinition of identities and presents ambivalent tensions between origin and host country affiliations \citep{balsamo2003famiglie}
Emotional networks may be strained by separation, yet transnational circulation of care and support can alleviate these challenges, facilitating ongoing connections across borders \citep{baldassar2014transnational}.

\subsection{Contribution of Albanian Women to Fertility in Italy}
 
According to Istat data, by 2001 the migrant population in Italy increased from 0.6\% to 2.3\% of the total population, with a notable feminization of migration flows that, coupled with the cultural diversity of migrants and their uneven spread across the country, has resulted in new and emerging migration patterns \citep{idos2022dossier}. From 2001 to 2009, the number of migrants tripled, reaching 6.5\%, surpassing four million in 2011 and five million in 2015, accounting for 8.3\% of the total population. By 2022, this percentage reached 8.7\%. Although the diversity of migrants' countries of origin, which characterized Italy's migration patterns since the 1980s and 1990s, remains, the relative proportions of each nationality have shifted significantly over time. As of January 1, 2022, immigrants from Romania, Albania, Morocco, China, and Ukraine made up 56\% of the total immigrant population and over half of the foreign female population in Italy.
Foreign female residents, who are in mean younger than Italian women, helped to mitigate the decline in births due to their substantial presence in the reproductive age group, their lower average age at childbirth, and a fertility rate close to the replacement level (1.89 children per woman in 2020). In 2022, approximately 13.5\% of births in Italy were to foreign-born parents, and 20.9\% to couples with at least one foreign-born partner \citep{istat2022}. 
The highest number of foreign children recorded in the national registry is Romanian (14,248 born in 2020), followed by Moroccan (9,991) and Albanian (8,082). These citizenships account for about 40\% of births to foreign mothers in Italy. 
However, The fertility rate of migrant women decreased significantly, with a sharper decline compared to natives (-18\% vs. -12\% since 2010) \citep{miaci2023}.
This convergence in fertility trends between migrants and natives is largely due to systemic barriers affecting reproductive choices. Italy's familistic welfare system offers limited public measures and services to balance work and parenthood, rendering migrant populations, particularly women, especially vulnerable. Migrant women often face precarious employment and have lower employment rates than their Italian counterparts (45.4\% vs. 49.9\% in 2012) \citep{OIL2023}. Additionally, migrant mothers report greater difficulties in balancing work and family \citep{bonizzoni2015uneven}, and are more likely to face challenges in accessing suitable childcare options \citep{mussino2023childcare, trappolini2023informal}, primarily due to language barriers and administrative complexities \citep{karoly2011early, 00fd45e84be54b85939b95cc228724e4}.
The recent study \cite{miaci2023}, based on the ISTAT birth registry data, shown that the fertility rates of Moroccan and Albanian women in Italy are particularly high (2.9 and 2.1, respectively) compared to both the native population (1.1) and other prevalent communities in Italy (1.5 for Romanians, 1.2 for Ukrainians, and 0.9 for Chinese). Moreover, the modal age at first childbirth is younger for Moroccans (around 25) and Albanians (around 26) compared to natives (about 32.8) and the average migrant population in Italy (29.7).
Furthermore, Albanian and Moroccan women exhibit higher TFRs in Italy than in their countries of origin, present an high risk of parity transitions and a significant arrival effect \citep{miaci2023, impicciatore2020migrants,mussino2012fertility}.
\cite{garcia2023comparison} found that the fertility intentions of Albanian migrant women are similar to those of Italian women and lower than those of non-migrants. \cite{impicciatore2020migrants} compared the reproductive patterns of Albanians, Moroccans, and Ukrainians with Italian non-migrants and stayers in the origin country. They found that the socialization hypothesis prevails for Moroccan women, while disruption explains the fertility of Ukrainian migrants. For Albanians, adaptation is the primary explanation. The risk of first childbirth among Albanian migrants decreases over time, converging to the level of Italian non-migrants about four years after arrival, a pattern not observed in other groups. 
On 2020, the gap in fertility rates between the entire migrant population and the Albanians and Moroccans are driven by compositional effects for the former and rate effects for the latter \cite{miaci2023}.
The fertility rate at the provincial level for Albanian women residing in Italy is relatively uniform across the territory, with a national rate of 2.1 children per woman, which is high in comparison to the rate in Albania (1.4 children per woman). 
Over the past decade, more than 83,000 children with Albanian citizenship born in Italy, representing about 15\% of all non-EU births in the country. The birth rate for the Albanian community in Italy remains higher than that of the overall non-EU population and the native population (15.7\% vs. 14\% and 6.5\%, respectively).

\begin{table}[H]
\centering
\caption{Total Fertility Rate, 2020 (ISTAT, UN)}
\begin{tabular}{|c|c|c|}
\hline
\textbf{TFR 2020 } & \textbf{Italy} & \textbf{Albania} \\
\hline
Italian & 1.2 & - \\
\hline
Albanian & 2.1 & 1.4 \\
\hline
overall migrant population & 1.8 & - \\
\hline
\end{tabular}
\end{table}

\subsection{Albanian community in Italy}

Albanian migration to Italy became notable from 1990 onwards, occurring in three major waves: the first in 1990-1991, a second in 1997, and a third during the Kosovo war in 1999, when many Albanians fled their country, often posing as Kosovars seeking asylum \citep{di2024effect}.

As of January 1, 2022, the Albanian community in Italy is well-established, with 396,918 legal residents, representing approximately 11\% of the non-EU population. The majority (60.5\%) reside in the northern regions, particularly in Lombardy, Emilia Romagna, and Tuscany. Rome also hosts a significant share (21\%) of the community. Although only 12.9\% live in the southern regions, there is a notable concentration in Puglia, with Bari accounting for about 5\% of the Albanian population \citep{ANPAL}.

\begin{figure}[H]
    \centering
    \caption{Percentage distribution of female Albanian population residing in Italian provinces.}
\label{fig: map}
    \includegraphics[width=1\textwidth]{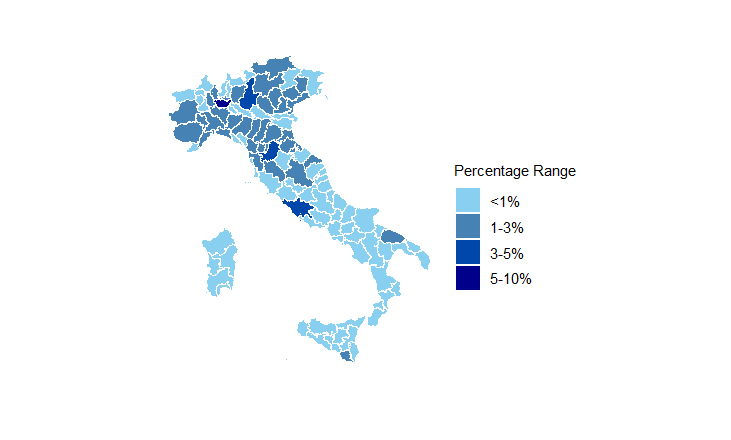}
\end{figure}

The community exhibits a balanced gender distribution (49\% women, 51\% men) and a younger age structure compared to the Italian population, with 42\% under 30 years old. There is a high presence of minors (25.2\%) and young women (30\% aged 18-35), while 13.2\% of the population is over 60 years old \citep{ANPAL}, consisting mainly of early migrants and the parents of first-generation migrants who later joined their children \citep{King2014}. 
Regarding residence permits, family reunification is the predominant reason for Albanian migrants \citep{ANPAL, toulemon2004fertility, gabrielli2019histories}. Long-term residents in 2020 make up 67\% of the Albanian community, a higher percentage than the overall non-EU population.
Recent studies of the Lombardy labor market indicate that Albanian workers experience better integration, with lower unemployment, higher employment rates, and less irregular employment compared to other migrants \citep {cela2022labour} and according to \cite{ANPAL}, the Albanian community in Italy shows slightly better employment outcomes compared to the overall non-EU population. The employment rate is 59.7\%, with a significant gender gap (77.3\% for men and 40\% for women).
For Albanian migrants, longer residence in Italy is associated with health deterioration, with migrants exhibiting poorer health than their co-nationals in Albania \citep{di2024effect}, indicating the disruptive effect of migration on health. This is consistent with the acculturation hypothesis: over time, migrants' health deteriorates, with recent migrants reporting better health than those with longer stays, aligning with the “exhausted migrant effect” \citep{bollini1995no}.

\begin{table}[H]
\centering
\caption{Residence permits issued to Albanian citizens in Italy, ISTAT 2022}
\begin{tabular}{|c|c|c|c|}
\hline
\textbf{Residence permits} & \textbf{Male} & \textbf{Female} & \textbf{Total} \\
\hline
\textbf{Family} & 10035 & 11039 & 21074 \\ 
 & 54.7\% & 67.9\% & 60.9\%\\
\hline
\textbf{Work} & 4342 & 1796 & 6138  \\
 & 23.7\% & 11.0\% & 17.7\% \\
\hline
\textbf{Study} & 159 & 264 & 423\\
 & 0.9\% & 1.6\% & 1.2\%\\
\hline
\textbf{Asylum} & 388 & 218 & 606 \\
 & 2.1\% & 1.3\% & 1.8\% \\
\hline
\textbf{Other} & 3412 & 2941 & 6353 \\
 & 18.6\% & 18.1\% & 18.4\% \\
\hline
\textbf{\textit{Total}} & \textit{18336} & \textit{16258} & \textit{34594} \\
\hline
\end{tabular}
\end{table}

\begin{figure}[H]
    \centering
    \caption{Residence permits issued to Female Albanian citizens in Italy}
    \label{fig:pie} \includegraphics[width=1\textwidth]{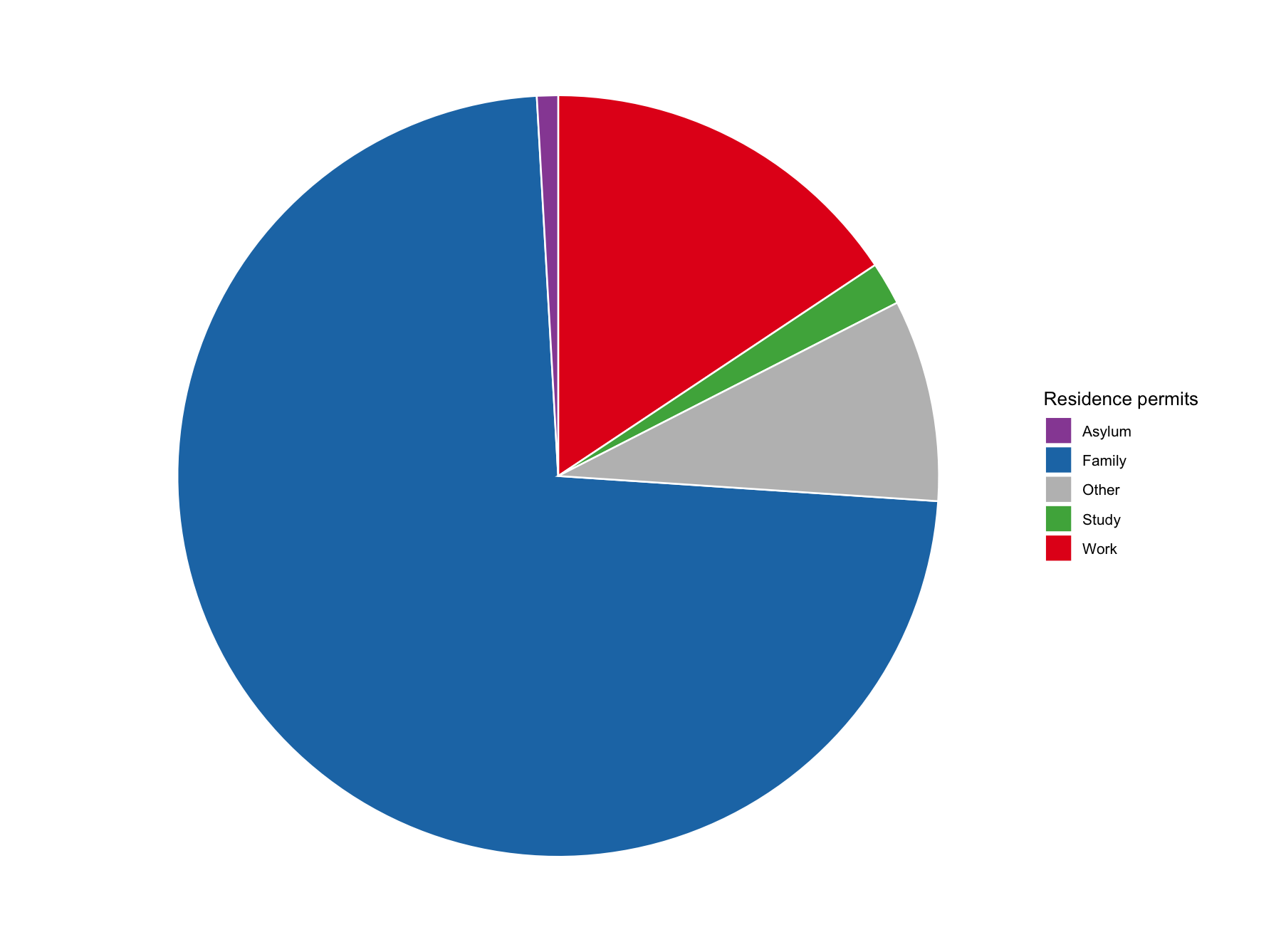}
\end{figure}

\section{Data}
\label{Sec:data}
Building on the macro data premises delineated in the background section, thirty semi-structured interviews were conducted to delve deeper into the expectations, representations, and practices related to motherhood and family life of Albanian women in Italy.
The textual corpus derived from the transcriptions of these interviews constitutes the research sample for this study. 

Given the focus on the maternity experiences of Albanian women in Italy, the interviews were conducted with women born in Albania, currently living in Italy, and mothers of children born or socialized in Italy (having migrated before the age of 6). The interviewees reside in the three most representative areas of Albanian presence in Italy, namely, the cities and surrounding provinces of Milan, Rome, and Bari.
Milan and Rome host a significant percentage of the total Albanian residents in Italy (42\% and 21\%, respectively). Bari, on the other hand, hosts a lower percentage of Albanian residents, about 5\%, but stands out compared to the rest of the South. In addition, Bari offers a unique context for studying how proximity migration can influence reproductive behaviors.

The characteristics of the interviewees, considered illustrative variables during data analysis, are summarized in Table \ref{tab:intervieweesCharac}.

\begin{table}[H]
    \centering
    \caption{Summary of the main variables collected during the interview, for each interviewee}
    \label{tab:intervieweesCharac}
    \begin{tabular}{ll}
        \hline
        \textbf{Variables} \\
        \hline
        Year of Birth (Age) & \\
        Place of Origin & \\
        Year of Arrival in Italy (Age at Arrival) & \\
        Entry Channel to Italy & \\
        Residence & \\
         Marital Status & \\
        Number of Children & \\
        Age of Oldest Child & \\
        Age of Youngest Child & \\
        Household & \\
        Level of education & \\
        Occupation & \\
        Use of external childcare & \\
        \hline
    \end{tabular}
\end{table}

For the selection of the sample in our study, we adopted a purposeful sampling strategy. The meetings took place in various environments, mainly at the interviewees' homes, centers where the interviewees volunteered, workplaces, and, in rare cases, on the Zoom platform. After presenting the study objectives and guaranteeing anonymity, each mother developed a biographical account of their motherhood and migration, supported by the interviewer's track.
The interview track is available upon request.
The main themes covered in the track are summarized in Table \ref{tab:mainthemes}.

\begin{table}[H]
    \centering
    \textbf{\caption{The main themes covered in the interviews}
    \label{tab:mainthemes}}
    \begin{tabular}{ll}
     \hline
        \textbf{Themes} \\
        \hline
       Pre-Migration Conditions in Albania \\
       Migration Journey \\
       Migratory Career \\
       Life in Italy and Integration Conditions\\
       Models and Imaginaries of Parenthood \\
       Family Construction and Division of Responsibilities \\
       Childcare and Support Networks\\
       Work and Maternity Reconciliation\\
       \hline
    \end{tabular}
\end{table}

The interviews, averaging 60 minutes in length, were recorded on audio tape and faithfully transcribed. Originally conducted in Italian, the shared language between the interviewees and the interviewer, the transcripts were later translated into English to make the results accessible to a wider, global audience. The interviews' transcriptions were edited to remove the interviewer’s words, leaving only the interviewees’ responses. All proper names and identifying information were replaced with pseudonyms before these texts became part of the study presented here.
The Key Features of the corpus are described in Table \ref{tab:KeyFeatures}.

\begin{table}[H]
    \centering
    \caption{Key Features of the Corpus}
    \label{tab:KeyFeatures}
    \begin{tabular}{ll}
     \hline
        \textbf{Key Features} \\
        \hline
        Documents & 30\\ 
        Tokens & 112,026\\
        Types& 4,155\\
         TTR & 3.7\%\\
        Hapax & 35.45\\
       Guiraud Index & 24.70\\  
       \hline
    \end{tabular}
\end{table}

\subsection{Preprocessing and data cleaning}

Preprocessing and data cleaning are fundamental steps in computational text analysis. These steps involve techniques for transforming raw texts into tokens (i.e., units of text, such as words or word stems) and methods for removing components that may introduce noise and hinder the extraction of key information \citep{welbers2017text}.

All analyses were conducted in R, with preprocessing and data cleaning carried out using the \texttt{quanteda} package \citep{quanteda}.
The steps taken were as follows:
\begin{itemize}
    \item Removal of punctuation, numbers, symbols, separators, and URLs. 
    \item Lemmatization: transforming tokens into their lemmas, which are the base or dictionary forms of words. 
    \item Removal of stopwords, i.e., words with little semantic value, such as articles, conjunctions, and fillers. 
    \item Removal of words with fewer than three characters. 
    \item Trimming: removal of infrequent words. We used a cutoff frequency of less than 20.
\end{itemize}

These steps in the analysis reduced our Document-Term Matrix (DTM) from 5,762 terms to 418.

\section{Methodology}
\label{Sec:method}

Latent Dirichlet Allocation (LDA) is a probabilistic model used for topic modeling that represents documents as mixtures of latent topics \citep{airoldi2015handbook}. Unlike traditional mixture models that assign each document to a single topic, LDA allows documents to be associated with multiple topics simultaneously.
Originally introduced by \cite{blei2003latent}, LDA is a Bayesian model for discrete data, operating under the assumption that topics are uncorrelated. In this framework, each document is linked with a set of topics, and the words within the document are generated based on the word distributions of these topics.
To further explain, conventional mixture models used in clustering assume that each observation (or document) originates from a single probability distribution. If an observation exhibits a high likelihood of belonging to multiple clusters, it reflects uncertainty in its assignment. In contrast, mixed membership models like LDA permit observations to belong to multiple clusters concurrently. This means that a document can encompass various topics, with each word potentially drawn from a different topic-specific distribution.
The primary objective of LDA is to develop a probabilistic model that not only assigns high probability to the documents within the corpus but also to other similar documents outside the corpus.

We adopt the notation presented in \cite{blei2003latent}, defined as:
\begin{itemize} 
\item A word is one of $V$ terms, indexed by ${1,\dots,V}$. 
\item A document is a sequence of $N$ words, denoted by $\bm{w} = (w_1, w_2, \dots, w_N)$, where $w_i$ represents the $i$th word in the sequence. 
\item A corpus is a collection of $M$ documents, expressed as $\bm{D} = {\bm{w}_1, \bm{w}_2, \dots, \bm{w}_M}$. \end{itemize}

The generative process of LDA can be described as follows:

\begin{enumerate} 
 \item For each topic, draw a term distribution $\beta \sim \mbox{Dirichlet}(\delta)$, which provides the probability of each word occurring within that topic. 
 \item For each document $w$: 
  \begin{enumerate} 
   \item Sample a topic distribution $\theta \sim \mbox{Dirichlet}(\alpha)$, representing the mixture of topics in the document. 
    \item For each word $w_i$ in the document: 
    \begin{itemize} 
    \item[a] Select a topic $z_i \sim \mbox{Multinomial}(\theta)$. \item[b] Choose a word $w_i$ from the conditional distribution $P(w_i|z_i, \beta)$, which is multinomially distributed based on the selected topic $z_i$.
   \end{itemize} 
   \end{enumerate} 
   \end{enumerate}

In this context, the dimensionality $k$ of the Dirichlet distributions (and hence the number of topics) is assumed to be known and fixed.
For model fitting, we used the R package \texttt{topicmodels} \citep{topicmodelsJSS}, which interfaces with implementations of the variational expectation-maximization (VEM) algorithm by \cite{blei2003latent} and the Gibbs sampling method by \cite{phan2008}. Our analysis specifically employed the VEM algorithm.
The VEM algorithm is an extension of the expectation-maximization (EM) algorithm \citep{dempster1977} tailored for Bayesian inference. While the EM algorithm focuses on maximum likelihood estimation to find the most probable parameter values, VEM approximates the entire posterior distributions of the parameters and latent variables. Like EM, VEM iteratively optimizes parameter estimates through an alternating procedure, but it operates within a variational framework to handle the complex dependencies and intractability issues that arise in fully Bayesian models.

LDA has been widely applied for topic modeling across various domains. \cite{jelodar2019latent} provide a comprehensive review of articles published between 2003 and 2016 that utilized LDA for topic modeling. More recent examples include \cite{Madzik2022}, who applied LDA to conduct a topic model analysis of a systematic literature review on the analytical hierarchy process, and \cite{garcia2023automatic}, who used LDA for topic modeling in financial news.

\section{Application and results}
\label{Sec:application}

To gain a preliminary understanding of the DTM, after preprocessing, an exploratory analysis was performed, then we employed the LDA model to delve into the main topics treated by each group of interviewees by entry channel.

\subsection{Exploratory analysis}

Visualizing the frequency and connection between words through word clouds, feature counts, and networks, allows to identify patterns and trends that might not be obvious from reading the text alone.
The insights gained from exploratory analysis can uncover potential areas of interest and guide into more in-depth analysis.

\begin{figure}[H]
    \centering
    \caption{Interviews word cloud.}
\label{fig: all_wordcloud}
    \includegraphics[width=1\textwidth]{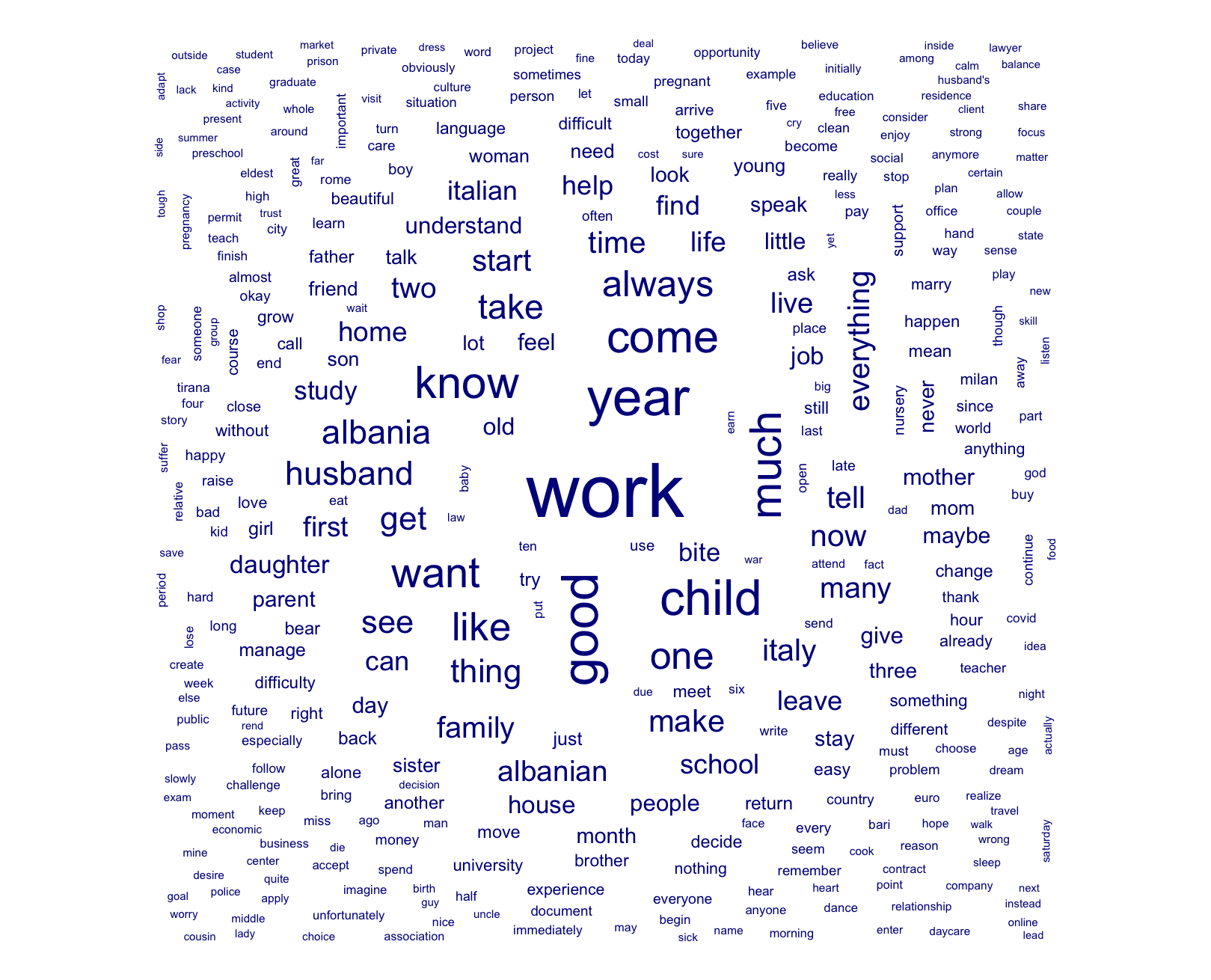}
\end{figure}

To identify the most significant words and themes at a glance, Table \ref{Tab:topwords} shows the most frequent 35 words and their counts. We also visually represent the most frequently occurring words in the corpus through a Word Cloud \ref{fig: all_wordcloud}, where the size of each word corresponds to its frequency, providing a quick and intuitive way to grasp the key terms and their prominence. 

\begin{table}[H]
\centering
\caption{Most used 35 words and their frequency.}
\label{Tab:topwords}
\begin{tabular}{|c|c|c|c|c|c|c|}
\hline
\textbf{work} & \textbf{year} & \textbf{good} & \textbf{child} & \textbf{come} & \textbf{know} & \textbf{much} \\
914 & 604 & 579 & 559 & 505 & 492 & 482 \\
\hline
\textbf{want} & \textbf{like} & \textbf{one} & \textbf{always} & \textbf{take} & \textbf{albania} & \textbf{husband} \\
454 & 417 & 403 & 374 & 351 & 338 & 333 \\
\hline
\textbf{get} & \textbf{thing} & \textbf{see} & \textbf{make} & \textbf{now} & \textbf{italy} & \textbf{tell} \\
330 & 326 & 321 & 303 & 296 & 295 & 294 \\
\hline
\textbf{everything} & \textbf{time} & \textbf{start} & \textbf{family} & \textbf{many} & \textbf{two} & \textbf{albanian} \\
293 & 292 & 291 & 283 & 277 & 273 & 273 \\
\hline
\textbf{live} & \textbf{first} & \textbf{life} & \textbf{home} & \textbf{school} & \textbf{can} & \textbf{leave} \\
267 & 258 & 256 & 254 & 252 & 248 & 247 \\
\hline
\end{tabular}
\end{table}

\begin{figure}[H]
    \centering
    \caption{Network of the top 30 words.}
\label{fig: network}
\includegraphics[width=1\textwidth]{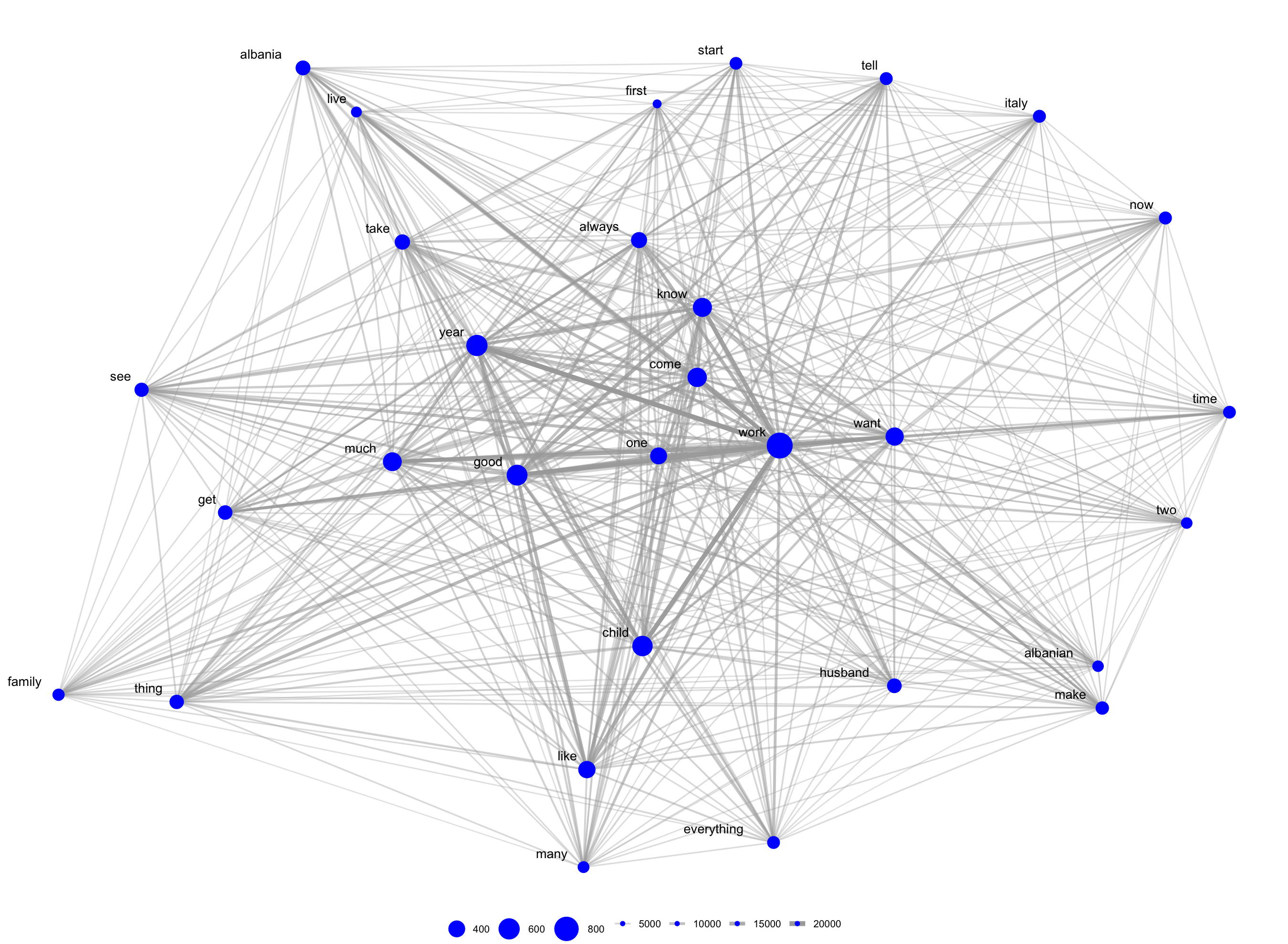}
\end{figure}

The network graph in Figure \ref{fig: network} of the most frequent 30 words, illustrate the key relationships between them. Each node represents one of the words, and the size of each node corresponds to the frequency of the word. Larger nodes indicate higher frequency. The edges between nodes represent connections between the words, indicating that the words frequently appear together within the text. The thickness of the edges denotes the strength of the connection, i.e., how often the words co-occur. Thicker edges suggest stronger co-occurrence relationships.
By examining the connections, we can understand the context in which words are used and how they are interconnected, indicating potential subtopics or themes within the corpus. Words that have many connections (highly interconnected nodes) might be central to the corpus, indicating their importance in multiple contexts.\\

The lexeme \textit{“work”}, which is the most frequent word, demonstrates also a connection with \textit{“start”} and \textit{“first”}, reflecting the beginning of careers or inaugural experiences. Moreover, its association with \textit{“Albania”} and \textit{“Albanian”} suggests occupational experiences in the country of origin and the presence of Albanian networks in Italy, which could be important in both professional and social spheres. The frequent co-occurrence of \textit{“work”} with \textit{“want”} hints a strong aspiration for employment, realised for some individuals and ongoing for others. Certain collocates (e.g., \textit{“much”}, \textit{“many”}, \textit{“always”}, \textit{“time”}) emphasise the amount of time and energy dedicated to occupation, while others attribute a positive connotation to it (\textit{“good”}), or emphasise the stability of their employment (\textit{“years”}). One of the most salient links is with the lexeme \textit{“child”}. A notable triangulation emerges between \textit{“child”}, \textit{“work”}, and \textit{“want”}, as well as between \textit{“child”}, \textit{“work”}, and \textit{“time”}. This pattern suggests that family and occupation are prominent, desired, and time-intensive domains for the interviewees.

The second dimension emerging as fundamental, alongside employment, appears to be the familial sphere. The lexeme \textit{“child”} features with high frequency. It is noteworthy that during the preprocessing phase, the decision was made to omit proper nouns, substantially diminishing the frequency of children’s names, which were often used in place of generic terms such as \textit{“child”} or \textit{“daughter/son”}. Additional high-frequency words include \textit{“husband”}, \textit{“family”}, and \textit{“home”}. An analysis of the patterns associated with \textit{“child”} reveals a frequent link with \textit{“want”}, often forming a triadic relationship with the numerals \textit{“one”} or \textit{“two”}. Other commonly associated words include \textit{“years”}, providing information on the offspring’s ages; \textit{“good”}, which may offer insight into behavioural aspects; and \textit{“Albanian”}, addressing matters of citizenship. The connection between \textit{“child”} and \textit{“husband”} could pertain to shared parental responsibilities and educational practices. A link is also observed between \textit{“child”} and \textit{“family”}, and between \textit{“child”} and \textit{“everything”}, alluding to the all-encompassing nature of family and motherhood.

Given that our primary research interest lies in understanding the experiences of motherhood among our interviewees and the impact of migration on their conceptualization of family, and considering that the literature identifies the entry channel as a significant predictor in this domain, we aggregated the interviews based on their respective entry channels.
Utilizing this aggregated corpus, consisting of three documents, as the 30 individual interviews have been grouped into three documents according to the reason for migration: labor and educational pursuits, family reunification, and irregular entry, we replicated the exploratory analyses.
The most frequent words and their links remain largely consistent with those derived from the disaggregated data, and for the LDA, we will display the results obtained from the aggregated corpus. However, to gain deeper insights into how Albanian immigrant mothers in Italy perceive and discuss the concept of “work” and “child”, a co-occurrence analysis was conducted on the interview transcripts. 

\subsection{Co-occurrence Analysis of “Work” and “Child”}

Co-occurrence analysis is a quantitative text mining technique that examines the frequency with which pairs of words appear together within a specified context, such as sentences or documents. This method allows for the identification of patterns and relationships between concepts in the text, providing insights into the underlying structure of the participants' discourse.

In the preprocessing stage, standard text processing techniques were applied, including tokenization, lemmatization, and stopword removal, to prepare the textual data for analysis. Notably, collocation detection was employed to identify multi-word expressions that frequently occur together, such as \emph{work-life balance} or \emph{take care}. This step is crucial as it preserves the semantic integrity of such phrases, ensuring that they are treated as single units in the analysis rather than separate words.

A Document-Term Matrix (DTM) was constructed using boolean weighting, where entries indicate the presence or absence of a term within a document (or sentence), rather than term frequency. This binary representation focuses on co-occurrence patterns without being influenced by the overall frequency of terms, which can skew the analysis toward more common words.

\begin{figure}[H]
    \centering
    \caption{Co-occurrence Network Graph Centered on \emph{work}}
    \label{fig:NETwork} \includegraphics[width=1\textwidth]{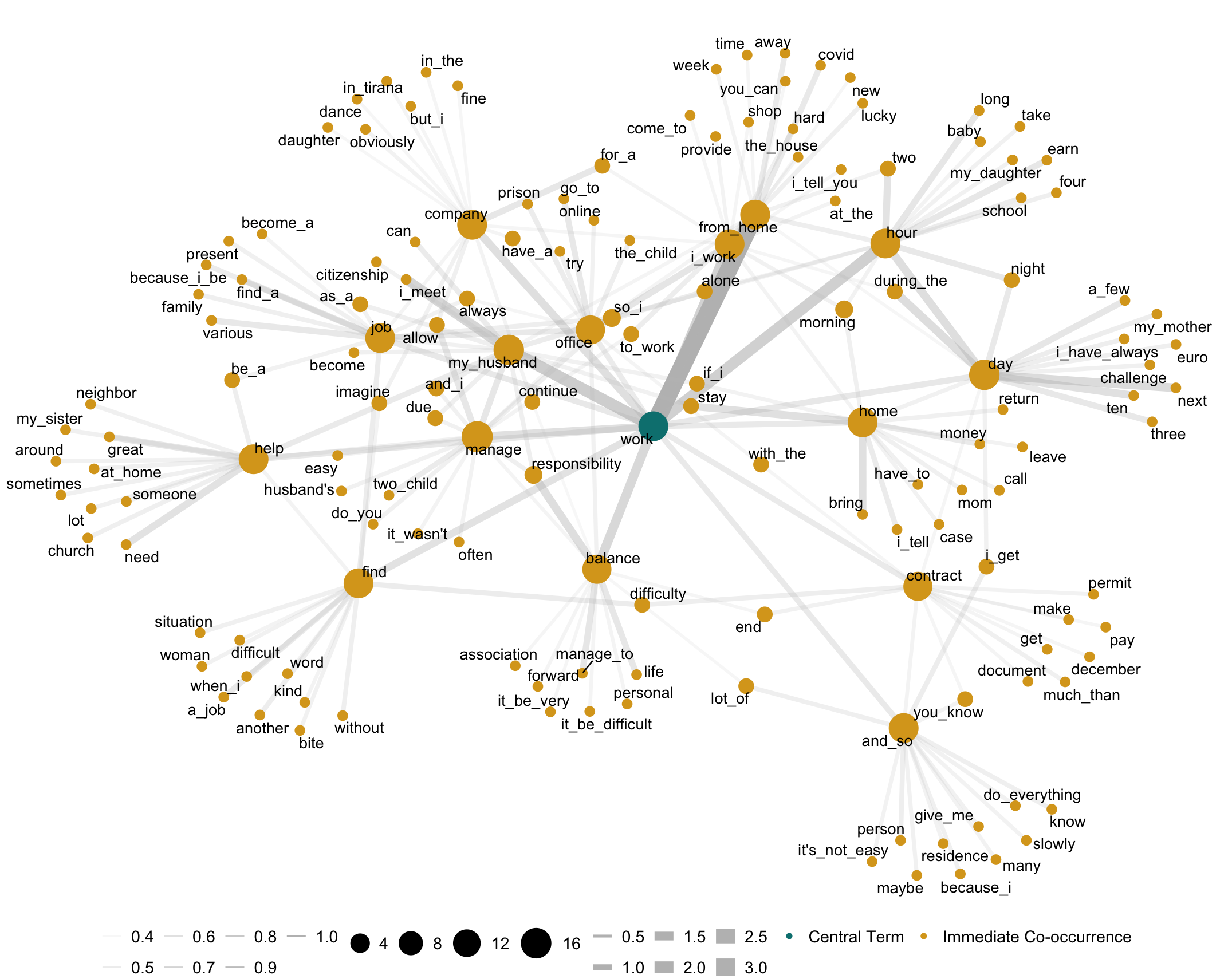}
\end{figure}

\begin{figure}[H]
    \centering
    \caption{Co-occurrence Network Graph Centered on \emph{child}}
    \label{fig:NETchild} \includegraphics[width=1\textwidth]{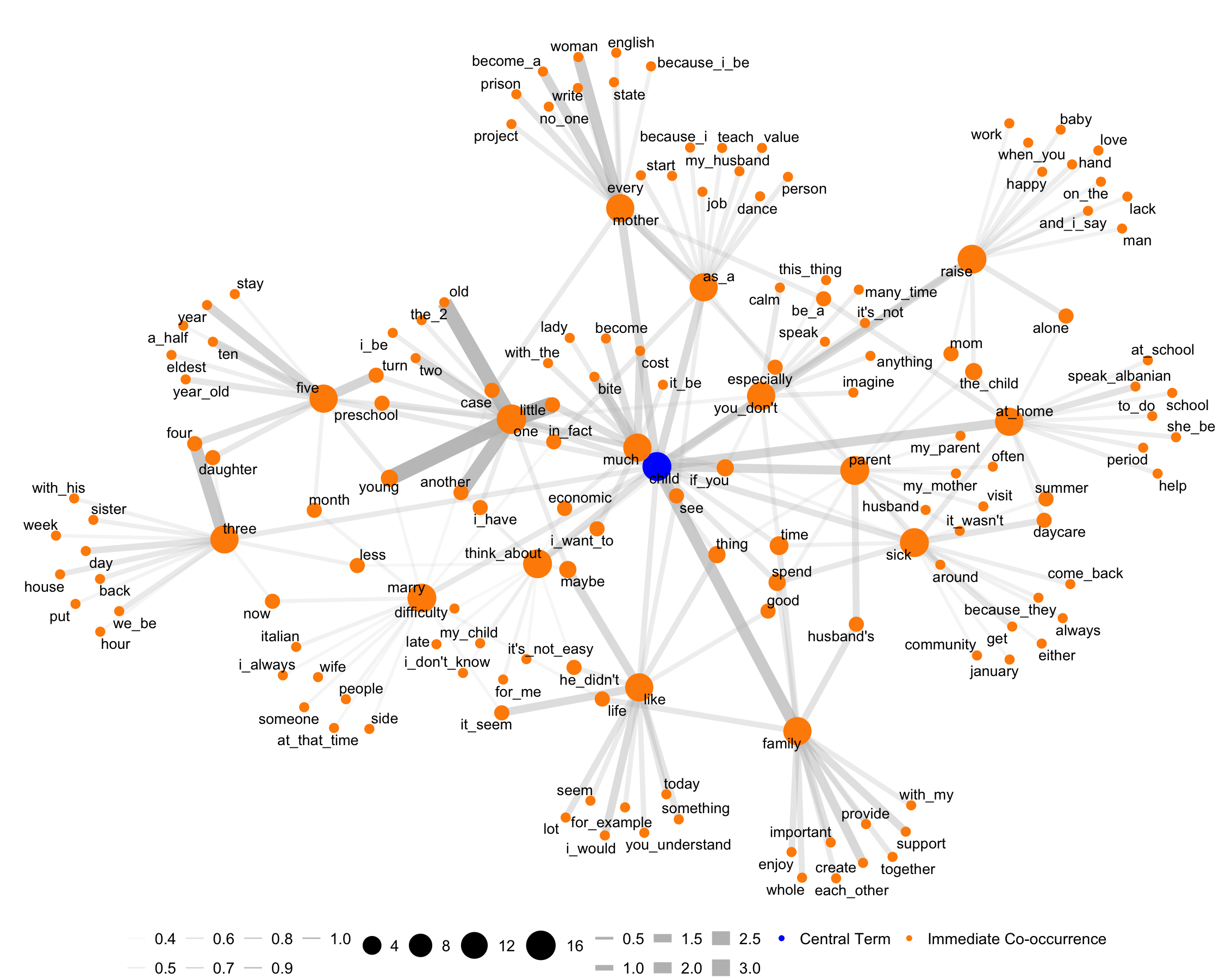}
\end{figure}

For the co-occurrence analysis, the terms \textbf{\emph{work}} and \textbf{\emph{child}} were selected as focal points. To quantify the strength and significance of their associations with other terms, the log-likelihood ratio (LLR) was employed \citep{manning2009introduction}. 
LLR assesses the statistical significance of the co-occurrence between two terms by comparing the observed co-occurrence frequency to what would be expected under the assumption of independence. The LLR is calculated using the formula:
\begin{equation*}
\begin{aligned}
\mbox{LLR} = 2 \bigg[ \, & M \cdot \log(M) - M_i \cdot \log(M_i) - M_j \cdot \log(M_j) + M_{ij} \cdot \log(M_{ij}) \\
& + (M - M_i - M_j + M_{ij}) \cdot \log(M - M_i - M_j + M_{ij}) \\
& + (M_i - M_{ij}) \cdot \log(M_i - M_{ij}) + (M_j - M_{ij}) \cdot \log(M_j - M_{ij}) \\
& - (M - M_i) \cdot \log(M - M_i) - (M - M_j) \cdot \log(M - M_j) \, \bigg],
\end{aligned}
\end{equation*}
where $M$ is the total number of documents, $M_{i}$ and $M_{j}$ are the document frequencies of terms $x$ and $y$ and $M_{ij}$ is the number of documents where both terms co-occur.
A higher LLR value indicates a more significant association, suggesting that the co-occurrence is unlikely to be due to chance \citep{dunning1994accurate}.\\
The most significant co-occurring terms with \emph{work} and \emph{child} are presented in Tables \ref{tab:COwork} and \ref{tab:COchild}, respectively.
To illustrate these associations, co-occurrence network graphs were generated for both terms (Figures \ref{fig:NETwork} and \ref{fig:NETchild}). In these graphs: Nodes represent terms, with the size of each node proportional to its degree centrality, indicating the number of direct connections it has with other terms. 
Edges represent the co-occurrence relationships between terms, with edge widths and opacities scaled according to the LLR values, reflecting the strength of the associations.
Node Colors differentiate between the central term (\emph{work} or \emph{child}) and co-occurring terms.

\begin{table}[h!] 
\centering 
\caption{Top Co-occurring Terms with \emph{work} Based on Log-Likelihood Ratio} 
\label{tab:COwork} 
\begin{tabular}{lr} 
\hline 
\textbf{Co-occurring Term} & \textbf{LLR} \\ \hline 
from home & 88.09 \\ 
my husband & 55.53 \\ 
hour & 52.25 \\
I work & 49.42 \\
balance & 42.33 \\
company & 34.68 \\ 
manage & 33.74 \\ 
find & 31.16 \\ 
help & 24.67 \\ 
day & 24.26 \\ 
\hline 
\end{tabular} 
\end{table}

For the term “\textit{work}”, the strongest association is with “\textit{from home}”, suggesting an adaptive strategy through which women attempt to reconcile professional and caregiving responsibilities. The link to “\textit{Covid}” in the graph stresses how the pandemic has intensified this trend.
The association with “\textit{my husband}” highlights the significant role of spouses, either as financial providers, or as influencers in women’s employment decisions, revealing the close ties between professional and family spheres.
The terms “\textit{balance}” and “\textit{manage}”, connected to “\textit{responsibilities}” and “\textit{difficulty}”, emphasise the struggle to achieve a sustainable work-life balance. “\textit{Balance}” is linked to words like “\textit{life}” and “\textit{personal}”, while “\textit{manage}” co-occur with “\textit{child}”, “\textit{husband}”, and “\textit{help}”, highlighting the interconnectedness of family responsibilities and the need for external support.
The link between “\textit{work}” and “\textit{help}” points to the crucial role of support networks. This may refer both to assistance in finding employment and managing childcare while working. Ramifications like “\textit{need}”, “\textit{someone}”, “\textit{neighbours}”, “\textit{sister}”, and “\textit{church}” in the graph suggest that this support often comes from informal networks within the community, reflecting a reliance on social ties. Lastly, the connection between “find” and “\textit{work}” illustrates the challenge many women face in securing employment, with the process of finding a job marked by both importance and difficulty. Notably, it is connected to “\textit{word}”, which may suggest that language barriers exacerbate this difficulty.
Another notable term co-occurring with “\textit{work}” is “\textit{contract}”, suggesting how employment intertwines with legal status concerns, adding another layer of complexity to women's professional decisions.

\begin{table}[h!] 
\centering 
\caption{Top Co-occurring Terms with \emph{child} Based on Log-Likelihood Ratio} 
\label{tab:COchild} 
\begin{tabular}{lr} \\
\hline 
\textbf{Co-occurring Term} & \textbf{LLR} \\ 
\hline 
family & 47.51 \\ 
at home & 36.31 \\
parent & 35.17 \\ 
mother & 33.05 \\ 
as a & 32.15 \\ 
raise & 32.08 \\ 
think about & 26.30 \\ 
sick & 24.08 \\ 
much & 23.99 \\ 
one & 23.22 \\ 
\hline 
\end{tabular} 
\end{table}

For the term \textit{“child”}, the strongest association is with \textit{“family”}, indicating that discussions about children are closely tied to broader family dynamics. This connection branches into terms of emotional (\textit{“love”}, \textit{“happy”}, \textit{“enjoy”}) and practical connection (\textit{“important”}, \textit{“provide”}, \textit{“each other”}, \textit{“support”}, \textit{“together”}).
The term \textit{“at home”} suggests that the home environment plays a significant role in conversations about children, possibly reflecting on childcare practices and the value of home life. Its associations with \textit{“speak Albanian”}, \textit{“alone”}, and \textit{“help”} further emphasise aspects related to education, cultural transmission and managing children within the home.
The associations between \textit{“child”} and \textit{“parent”} or \textit{“mother”} underline the centrality of parental figures in children's lives. \textit{“Parent”} also refers to the importance of participants’ own parents, as shown by the ramifications \textit{“my parent”}, \textit{“my mother”}, \textit{“husband's”}, \textit{“often”}, and \textit{“visit”}.
The term \textit{“as a”} may refer to expressions like "as a mother" or "as a parent," indicating self-reflection on parental roles and responsibilities. \textit{“Raise”} highlights discussions about child-rearing, with associations like \textit{“when you”}, \textit{“work”}, emphasising the challenge of balancing work and parenting, and \textit{“alone”}, \textit{“lack”} indicating the difficulty of missing a support network. Terms like \textit{“love”} and \textit{“happy”} point to the emotional dimensions of raising children.
\textit{“Think about”} may reflect considerations around parenting decisions, such as planning for a child's future or contemplating having more children. This last interpretation is suggested by ramifications to \textit{“having another”}, \textit{“I don't know”}, \textit{“it's not easy”}, \textit{“late”}, and \textit{“difficulty”}.
The term \textit{“sick”} highlights concerns related to children’s health, particularly the challenges of managing childcare when a child is unwell. Its association with \textit{“at home”}, \textit{“always”}, and \textit{“daycare”} reflects how daycare is viewed as a partial solution to the challenges of work-life balance.
Lastly, the co-occurrence with \textit{“much”}, linked to \textit{“cost”} and \textit{“time”}, suggests considerations about the demands of parenting, particularly in terms of resource and time allocation.

\subsection{Latent Dirichlet Allocation application}

For the LDA model the number of topics $k$ has to be fixed a-priori. We run the model for $k$ ranging from 2 ro 10 and selected the best parametrization using the Bayesian information criterion (BIC). 

\begin{figure}[H]
    \centering
    \caption{BIC values for the LDA model with the number of topics $k$ ranging from 2 to 10.}
\label{fig: LDA_bic}
    \includegraphics[width=1\textwidth]{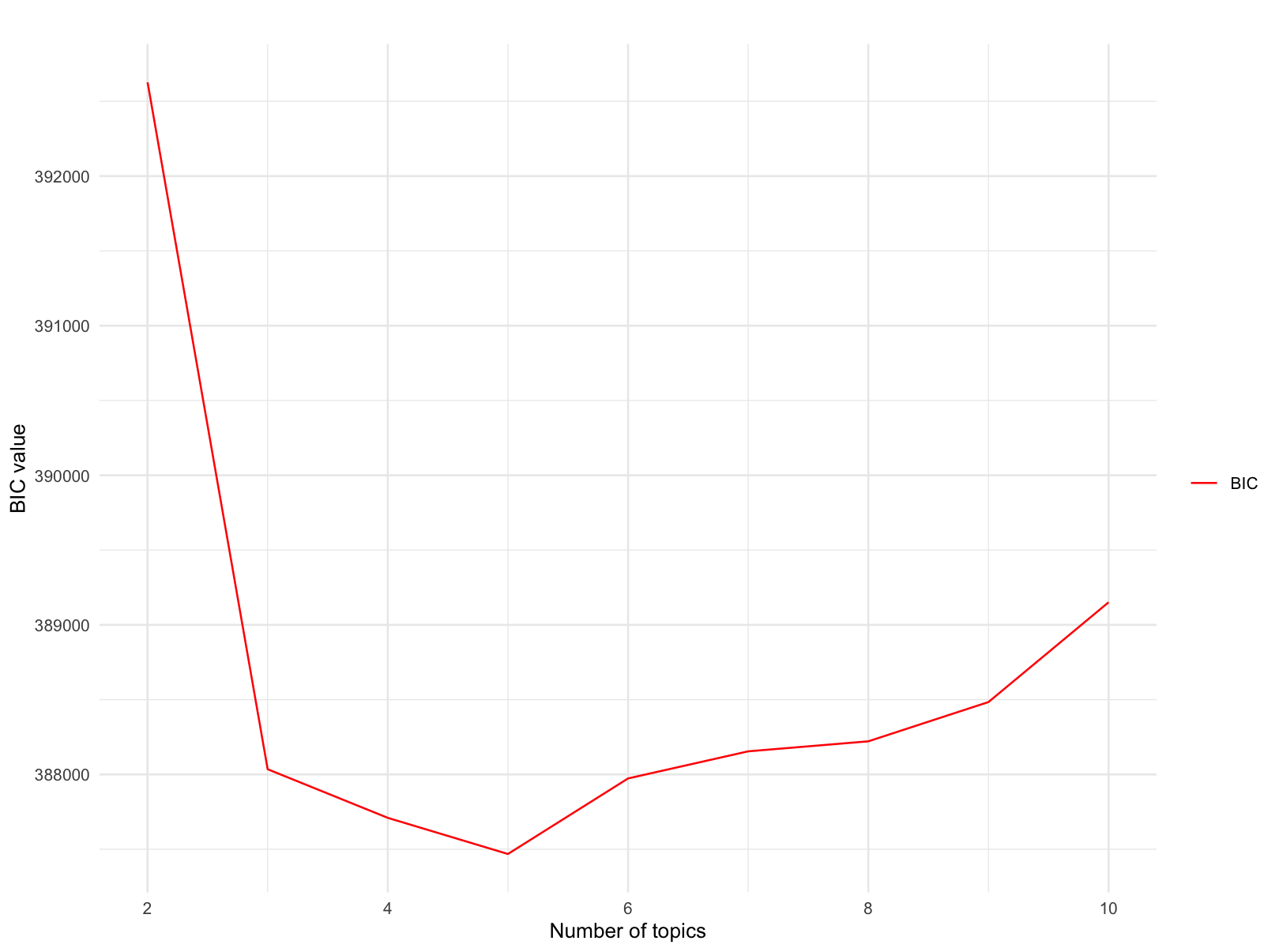}
\end{figure}

As shown in Figure \ref{fig: LDA_bic}, the lowest BIC is for 5 topics. However, as highlighted by \cite{FraleyHowManyClusters}, one of the main parameters when choosing the number clusters is the interpretability. We obtain close BIC values for 4 and 5 clusters, but the results are much more interpretable with a more parsimonious number of groups. This is why we decided to select 4 topics.

\begin{figure}[H]
    \centering
    \caption{Subcorpora identified by stratifying the sample based on the “reason for migration” variable (irregular, reunification, and study and work channels), modeled as a mixture of the four identified topics.}
\label{fig: LDA_channel}
    \includegraphics[width=1\textwidth]{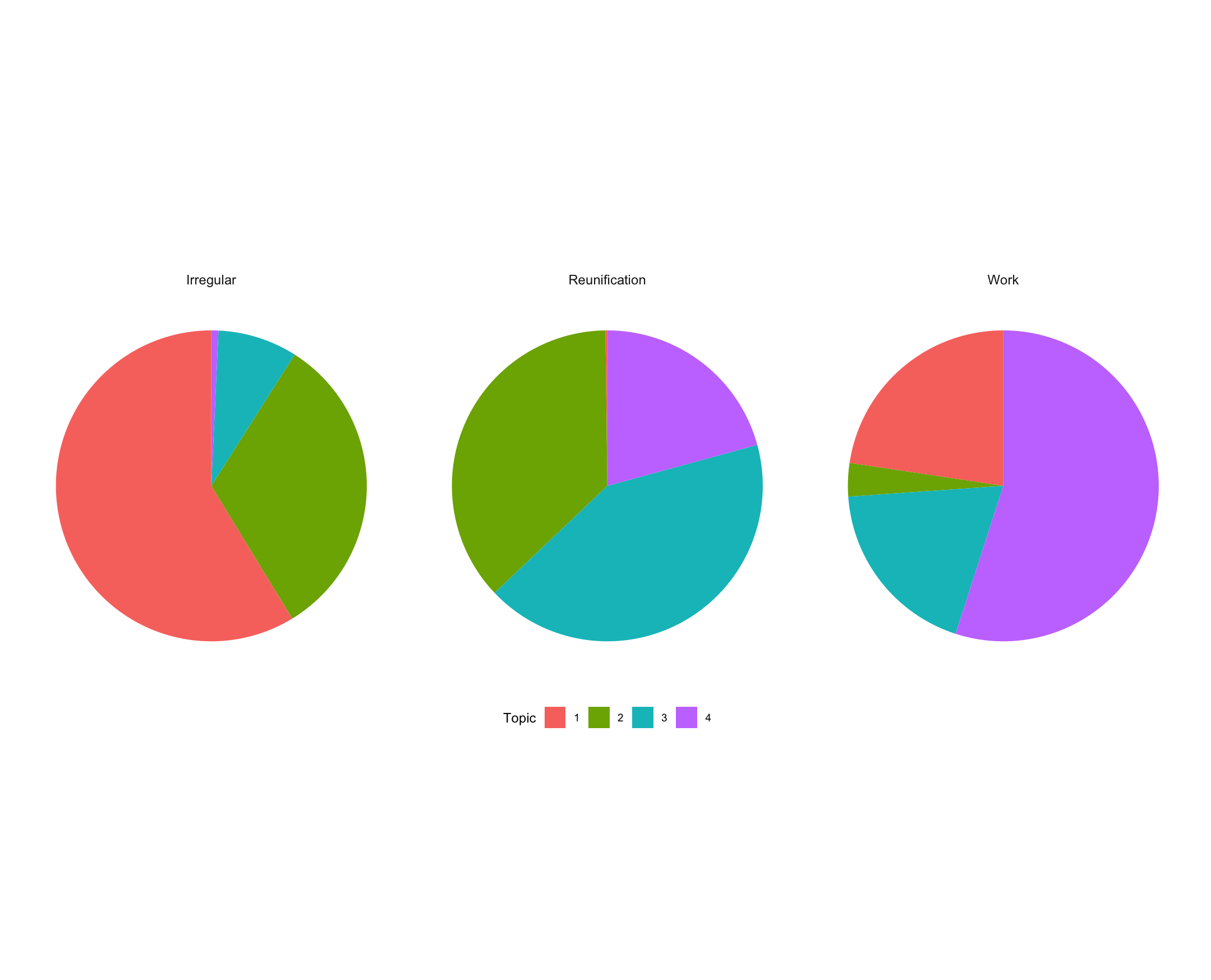}
\end{figure}

The LDA clarified how these topics model each subcorpus (Figure \ref{fig: LDA_channel}).
We determined that the most comprehensive interpretative approach was to analyze the topics along with their subcorpora modeling. 
Based on the lexical composition of each topic, we conducted an interpretative analysis to assess both the thematic content and its relative relevance within each subcorpus, a process that inherently involved a degree of subjectivity.
Given that the interviews were conducted within a predetermined framework and aimed to explore the thematic domain of motherhood, as highlighted in our exploratory analysis, we acknowledge the presence of thematic overlap in areas such as familiar responsibilities and work-life balance. However, certain lexemes exhibit differential semantic and contextual significance between topics.

\begin{figure}[H]
    \centering
    \caption{Comparison of terms across the four topics. The x-axis lists the terms, while the y-axis represents the estimated probabilities of their association with each of the four topics.}
\label{fig: LDA_words}
    \includegraphics[width=1\textwidth]{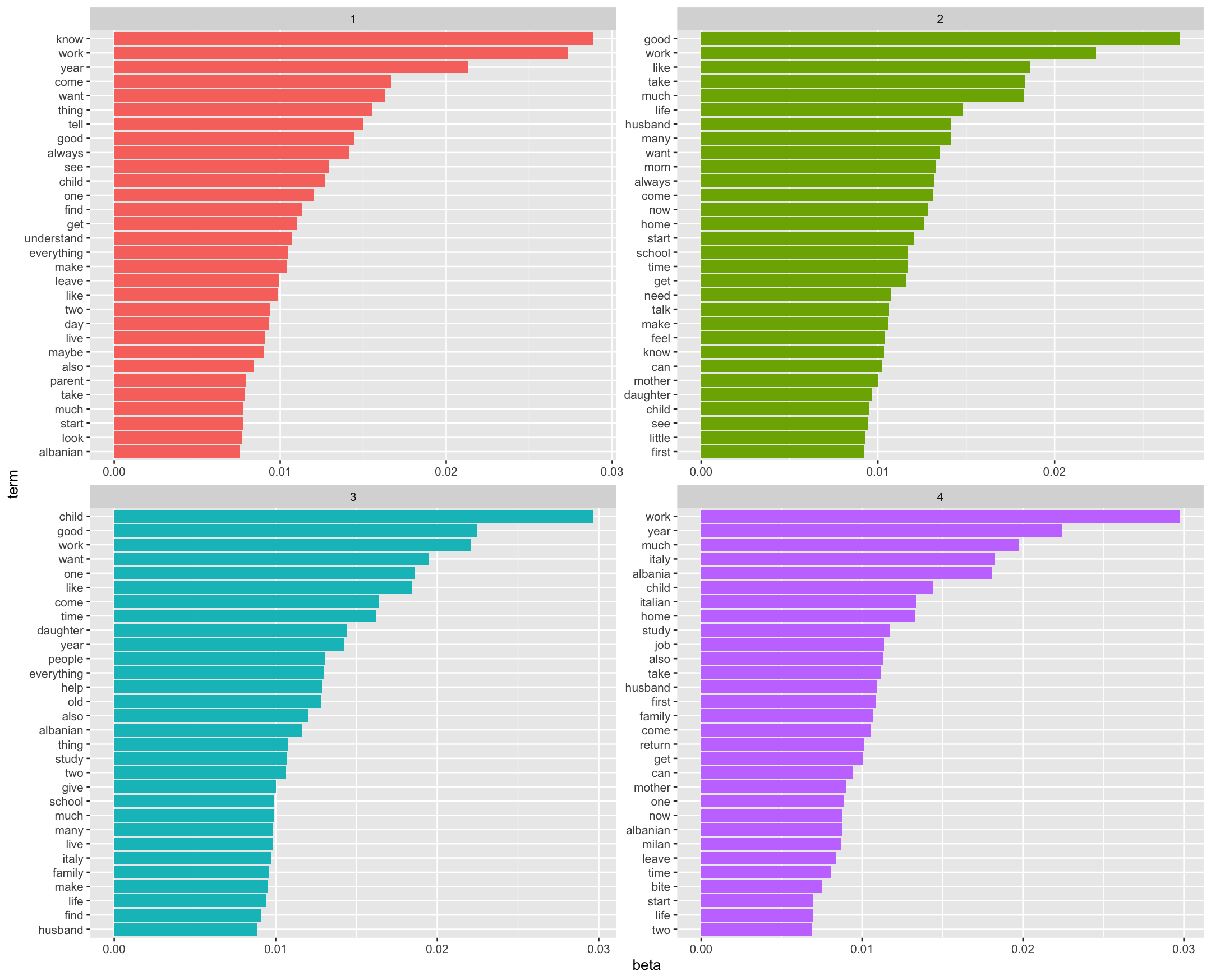}
\end{figure}

\begin{enumerate} 
\item We labeled the first topic “\textbf{Settlement}” due to preponderance of verbs associated with acclimatization and orientation processes. The key lexemes include: ``\textit{know}," ``\textit{come}," ``\textit{want}," ``\textit{tell}," ``\textit{see}," ``\textit{find}," ``\textit{get}," ``\textit{understand}," ``\textit{make}," ``\textit{leave}," ``\textit{live}," and the adverb of probability ``\textit{maybe}." In contrast to other clusters where work and family-related terms predominates, this topic emphasizes the initial phases of environmental adaptation.
This topic exhibits particular salience within the subcorpus of migrants who arrived under irregular conditions and maintains a notable presence in the subcorpus of migrants who immigrated for occupational or educational purposes. However, its representation in the subcorpus of women who migrated for reunification purposes is minimal, constituting less than 5\% of the content.
The minor prominence of the “Settlement” topic within the reunification subcorpus may be attributed to the pre-existence of a support network in Italy, typically comprising friends or family members, which facilitates integration. This established network smooths the acclimatization process, providing immediate social and emotional support, guidance, and resources that mitigate the stressors associated with relocation to a foreign country. Conversely, migrants arriving for occupational or educational purposes are often primary migrants lacking such established support systems \citep{de2010internal}. Consequently, they must navigate the complexities of a new environment autonomously, and the process of understanding cultural, social, and logistical landscapes may be more pronounced for them.
Migrants who entered the country irregularly encounter additional obstacles. They often lack legal protections and access to formal employment and housing, exacerbating their initial settlement difficulties. The imperative to secure employment, obtain accommodation, and comprehend legal and social systems under precarious conditions is more pressing for this group. The amplified challenges faced by irregular migrants are evident in their need to navigate survival strategies, often within the informal economy or through temporary, unstable arrangements \citep{Engbersen2009The}. This precarious situation renders the settlement phase particularly challenging, necessitating constant adaptation to a volatile environment. The presence of the modal adverb ``\textit{maybe}" underscores the uncertainty and tentative nature of the settlement experience. The decision-making process is frequently fraught with unknowns, and plans may undergo modifications based on evolving circumstances.

\item The second identified topic has been labeled as ``\textbf{A New Start}," characterized by lexemes reflective of planning and aspirations. Key verbs include: ``\textit{like}," ``\textit{want}," ``\textit{need}," ``\textit{take}," and ``\textit{feel}." Important nouns are: ``\textit{work}," ``\textit{home}," ``\textit{start}," and ``\textit{time}." Significant adverbs include: ``\textit{good}" and ``\textit{now}."  These terms collectively underscores themes of renewal and forward momentum.
The prominence of this topic among migrants arriving under irregular circumstances and those who migrated for reunification purposes suggests that, subsequent to initial settlement challenges, there is shift towards planning and building a better future. For woman who arrived without legal status, this transition may represent an evolution from immediate survival strategies to longer-term plans for stability and socioeconomic advancement. This group may now be prioritizing the finding of stable employment, the establishment of more permanent home, and integration into the host community.
With respect to migrants who immigrated for reunification purposes, we interpret the relevance of this topic as indicative of a potential reorientation towards personal objectives and aspirations that may have been previously subordinated to familial obligations. The emphasis on ``\textit{work}," ``\textit{time}," and ``\textit{home}" suggests that these individuals are seeking to balance family responsibilities with personal and professional growth, aspiring for economic stability and social integration. The words ``\textit{good}" and ``\textit{now}" reflect an optimistic perspective and a sense of immediacy. This optimism may be attributed to a combination of personal resilience and perceived environmental opportunities. This transitional phase in the migratory experience, characterized by a shift from reactive to proactive strategies may be influenced by factors such as increased familiarity with the host country's social and economic systems, improved language proficiency, and the development of social networks.

\item The third topic is titled ``\textbf{Family Life}” as it is characterized more than the other topics by a higher frequency of words related to family and domestic life. The most prevalent lexeme is ``\textit{child}", accompanied by other words such as ``\textit{daughter}", ``\textit{family}", ``\textit{husband}", and ``\textit{old}". The recurrent use of ``\textit{everything}” in association with ``\textit{family}” underscores the importance ascribed to familial relationships by respondents. Lexemes such as ``\textit{help}”, ``\textit{need}”, ``\textit{time}”, ``\textit{school}”, and ``\textit{work}” are prominent, reflecting the pragmatic aspects of familial management and the complex interplay between familial responsibilities and external obligations. The inclusion of ``\textit{old}" may indicate considerations of intergenerational dynamics.
This topic covers the largest area in the subcorpus of mothers who immigrated to Italy for reunification purposes, aligning with the literature that identifies  familial responsibilities and caregiving as priority for this group \citep{ortensi2015engendering}. Women who migrated for occupational or educational purposes , as well as under irregular circumstances, also demonstrate significant representation within this topic, reflecting their maternal roles and indicating the pervasive importance of family life across diverse migrant categories.

\item The fourth identified topic has been labeled “\textbf{Employment and Cross-Cultural Dynamics}” due to its encompassing words related to two significant aspects. Firstly, the word ``\textit{work}” has the highest estimated probability of association in this topic and there is a higher prominence of lexemes related to employment and study, such as ``\textit{job}”, ``\textit{study}”.
Moreover, their migratory trajectories to Italy, potential considerations of return migration to Albania, and the maintenance of transnational ties with their country of origin (e.g., ``\textit{Italy}”, ``\textit{Albania}”, ``\textit{Italian}”, and verbs such as ``\textit{return}”, ``\textit{leave}”). 
This topic is most prominent in the subcorpus of migrants who relocated for study or work reasons and it also partially explains the subcorpus of migrants who came for reunification purposes.
Migrants who came to Italy for educational or occupational purposes are likely to be deeply engaged in securing employment or educational placements, navigating the Italian labor market, or pursuing their studies. The high frequency of words such as ``\textit{work}”, ``\textit{job}”, ``\textit{study}” highlights their primary focus on professional and academic development. The specific mention of ``\textit{Milan}” might be attributed to the fact that our interviewees who came for work and study reasons are primarily residents in Milan. The importance of urban centers as hubs for work and study migrants is well-established, and Milan, being a major economic and educational nexus in Italy, attracts a substantial number of these migrants, reflecting broader migration patterns captured in ISTAT macrodata.
This topic also partially applies to those who migrated for reunification purposes. For these migrants, integrating into the workforce or pursuing further education may be key aspects of their integration and long-term settlement plans. The diminished relevance of this topic for former irregular migrants reflects how their undocumented status inhibited their access to formal employment and education opportunities.
The recurrence of lexemes such as ``\textit{Italy}", ``\textit{Italian}", ``\textit{Albania}" and ``\textit{Albanian}" suggests that migration often involves maintaining transnational connections with the country of origin \citep{baldassar2013locating}. This might encompass frequent visits, communication with family back home, remittances, or plans for return migration. 
The minimal explanatory power of this topic for migrants who arrived without legal status may be attributed to their initial restricted mobility options due to their former undocumented status, to their potentially worst economic condition and their prioritisation towards integration in Italy \citep{gabriel2008changing}. 

\end{enumerate}

\section{Discussions and Conclusions}
\label{Sec:conclusion}

This study investigated the motherhood trajectories of Albanian women in Italy, shedding light on the experiences of one of the country's most established migrant communities. While previous research has addressed various aspects of Albanian migrants' lives in Italy, our study fills an important gap by focusing on maternal practices and experiences, a dimension that has been underexplored.

Using a quantitative analysis of interview material through statistical text analysis, we delved into the thematic structures of these narratives. The validity of this approach in analyzing sociologically relevant corpora is well-established \citep{bolasco1999paradigmatic}, with several studies demonstrating its effectiveness in uncovering deeper insights from interview data \citep{shrader2021collusion, hewitt2021leadership, chandler2008academics, migliore2006children}. 
 
Textual analysis strikes a balance between rigorous analysis and interpretative contributions from the researcher, reducing uncertainties commonly associated with qualitative approaches, while maintaining a reasonable level of intersubjectivity and reproducibility. Furthermore, our use of Latent Dirichlet Allocation (LDA) enabled a nuanced exploration of the interviewees' complex and multi-dimensional experiences.

The exploratory analysis revealed two primary dimensions dominating the narratives of Albanian mothers: family and work. The network of the most used words, highlighted that most words were connected to the larger nodes of ``child" and ``work", with ``time" emerging as a critical connecting point. This triangulation reflects the significant time and energy demands that both family responsibilities and work obligations place on these women, underscoring the challenges of achieving work-family balance.
The co-occurrence analysis allowed us to gain a deeper understanding of the dynamics linked to these two domains, revealing interesting connections to informal support networks, concerns about children's upbringing and cultural transmission, and strategies for balancing work and family through remote work and flexible schedules.
 Our findings align with previous research that highlights the multiple disadvantages faced by migrant women in Italy, including employment challenges, economic and social deprivation, limited support networks, and restricted access to welfare benefits and childcare\citep{karoly2011early, bonizzoni2015here, paterno2016immigrants}. 

The LDA analysis, stratified by entry channels, revealed four main topics: ``Settlement," ``A New Start," ``Family Life," and ``Employment and Cross-Cultural Dynamics".

 Our findings align with prior research \citep{kulu2005migration, nedoluzhko2007migration, mussino2012fertility, ortensi2015engendering} in underlining that motivations for migration may significantly influence fertility decisions and life paths. Women who migrated for family reunification tended, in their narratives, to primarily focus on family dynamics and household management, while those migrating for employment emphasised work-related challenges and cross-cultural encounters.
The topic of ``Settlement" captured the complexity of environmental adaptation, particularly for migrants who arrived under irregular conditions or for employment and educational opportunities. Family reunification migrants, on the other hand, may have experienced a smoother initial transition due to pre-existing support networks. Some migrants with a history of irregular status expressed a strong desire to secure a better future, indicating resilience in overcoming early challenges. Those migrating for family reunification, meanwhile, often described ambitions for personal and professional growth, suggesting aspirations for self-actualization beyond familial roles.

One limitation of this study is that, despite covering the main settlement areas of the Albanian population in Italy, the small sample size of interviews limits the generalizability of our findings to the broader Albanian population in the country. Future research would benefit from expanding the sample of interviewees to strengthen the representativeness of the findings.

In conclusion, this study makes a significant contribution to understanding the intricate relationship between migration, motherhood, and integration, particularly in the context of Albanian women in Italy. By highlighting the diverse experiences shaped by different migration pathways, this research offers valuable insights for policymakers and practitioners seeking to better support migrant families.

\section{Acknowledgements}

This study is realized within the framework of the project “Multiple Ethnic Inequalities: A Multidimensional Analysis of the Penalty for Migrants in Occupational Achievement and Health in Italy (MEI)” led by the Department of Social and Political Sciences of the University of Milan.

The research protocol was reviewed and approved by the Ethics Committee of the University of Milan, with the ethics approval reference number 3/24, issued on January 16, 2024.

\bibliographystyle{tfq}
\bibliography{mrad}

\end{document}